\documentclass{article}
\usepackage[utf8]{inputenc}
\usepackage{authblk}
\usepackage{amsmath}

\begin{document}

\title{Non-linear interactions in cosmologies with energy exchange}
\author{John D. Barrow$^{1}$and Georgia Kittou$^{2}$ \\
$^{1}$DAMTP, Centre for Mathematical Sciences,\\
University of Cambridge\thanks{J.D.Barrow@damtp.cam.ac.uk}\\
Wilberforce Rd., Cambridge CB3 0WA, UK\\
$^{2}$College of Engineering and Technology, \\
American University of the Middle East, Kuwait\thanks{georgia.kittou@aum.edu.kw}}
\maketitle

\begin{abstract}
We investigate the case of two interacting fluids in homogeneous and
isotropic cosmologies with a non-linear interaction term. The interaction
term avoids the unrealistic form generally used in the literature, beginning
with Tolman, in which the interaction is zero when the Hubble parameter
vanishes. A variety of exact solutions for the scale factor are found and
describe a range of new behaviors. We also extend the analysis of possible
cosmological solutions with interacting fluids when curvature is taken into
consideration. We use an example of energy exchange between radiation and
scalar field to alleviate the flatness problem.
\end{abstract}

\section{Introduction}

In this study, we consider cosmological models that can describe the
observed acceleration of the universe, by modifying the standard picture of
the material content of the universe. We introduce a multi-fluid universe
with a mutual energy exchange between the fluid components, in such a way
that the total energy is conserved \cite{51,15,16}. Assuming that one of the
fluids belongs to the 'dark sector', then the cosmic acceleration can be
caused by the late dominance of this fluid.

The study of energy exchange between cosmological fluids is quite extensive
because of its wide applicability. Many physical processes can be modelled
by interacting fluids: particle-anti particle annihilation into radiation 
\cite{page, 80}, the process of black-hole formation and quantum evaporation 
\cite{9, kolb, barLid}, and the decay of other forms of mass-energy into
radiation. A list of specific interaction functions can be found in
literature, see \cite{barrow-clifton1}. The interaction function where
energy exchange is proportional to the product of the Hubble expansion rate
times an arbitrary linear combination of energy densities of the two fluids
appears by Barrow and Clifton in \cite{barrow-clifton1} and encompasses many
of these applications. An interaction term proportional to the product of
energy densities of dark matter and dark energy was introduced in Ref.\cite%
{27,inhomogeneous}. This interaction function is widely used in ecology to
describe prey-predator models. In cosmology such an interaction term is used
as a way-out\ of the coincidence problem \cite{53,90}. A more general ansatz
for the interaction is used in the so-called Chaplygin gas model \cite%
{inhomogeneous} to present a unified dark matter and dark energy model with
the first fluid dominating at early times and the other dominating at late
times. By introducing an energy exchange between scalar and matter fields in
scalar tensor theories, one can also examine the erosion of the value of the
gravitational `constant' $G$ over cosmological timescales.

When non-zero curvature is included in the field equations, the behavior of
the scale factor changes significantly at large expansion times. For
positively curved universes, oscillatory solutions can be obtained. In cases
of spatial flatness, it was shown in \cite{28} that oscillatory behavior can
be produced when one of the fluid components has a negative (`ghost') vacuum
energy. In ref. \cite{mag} a fuller description shows how exact solutions
for energy exchange can describe oscillating closed universes with constant
and increasing entropy in cosmologies with both varying, and unchanging
physical, constants.

The plan of the paper is as follows. In Section \ref{Section2} we introduce
a new interaction term where the Hubble parameter is not contained
explicitly and explain why this is physically more realistic than previous
models. In Section \ref{Section3} we write down the field equations and
arrive in an autonomous master equation related to the Hubble expansion.
Then, two different approaches are followed to obtain exact solutions for
the expansion of the scale factor. In Section \ref{Section4} the effects of
spatial curvature are included in the field equations and in the last
section we summarise our results. 

\section{The interaction term}

\label{Section2} Consider a flat Friedmann-Robertson-Walker (FRW) universe
containing two cosmological fluids $\rho _{1},\rho _{2}$ that are exchanging
energy in such a way that the total energy is conserved. We introduce an
interaction term, $Q$, so that the individual equations of energy
conservation for both fluids are given respectively by 
\begin{eqnarray}
\dot{\rho _{1}}+3H\Gamma \rho _{1} &=&Q,  \label{rho_2} \\
\dot{\rho _{2}}+3H\gamma \rho _{2} &=&-Q,  \label{rho_2a}
\end{eqnarray}

\noindent and the total density is conserved. Here, $\Gamma $ and $\gamma $
are the barotropic indices for each fluid,

\begin{equation*}
p_{1}=(\Gamma -1)\rho _{1}\;\text{and \ }p_{2}=(\gamma -1)\rho _{2},
\end{equation*}

and $H=\dot{a}/a$ is the Hubble parameter; $a(t)$ expansion scale factor of
the universe and $t$ is comoving proper time.

An energy-conserving exchange model of this sort was first introduced by
Tolman \cite{Tolman} in 1934 to describe the transfer of energy from matter (%
$\rho _{2}$) to blackbody radiation ($\rho _{1}$) in the universe -- `from
the nebulae into internebular space' as he described it. Tolman chose $%
Q=\gamma H\rho _{2}$, where $\gamma >0$ is a (small) constant ($\gamma
\simeq 10^{-7}-10^{-4}$), so a first integral is found to be $\rho
_{2}\propto a^{-3-\gamma }$\footnote{%
In \cite{Tolman}, Tolman's $g(t)=2\ln (a)$ and his mass function is $%
M(t)=\frac{4\pi}{3}\rho _{m}a^{3}.$}.

From a dimensional point of view it is expected that the interaction term, $%
Q $, should be a function of the Hubble parameter and energy density \cite%
{Tolman, 9, inte2}. In \cite{barrow-clifton1}, the authors considered an
interaction term of the form $Q=H(-\beta \rho _{1}+\alpha \rho _{2})$ to
describe a mutual energy exchange between two cosmological fluids. Their
choice complies with the current observational data. In particular, to
alleviate the coincidence problem \cite{coi1,coi2}, an interaction between
dark matter and dark energy is taken into consideration, and it is shown
that for such interaction term the energy densities $\rho _{1}$ and $\rho
_{2}$ evolve at the same rate.

Almost all past studies of interacting fluids, from Tolman onwards, have
assumed $Q\propto H\rho $. This is an obvious mathematical reason to do
this: it allows for an immediate integration of (\ref{rho_2}) or (\ref%
{rho_2a}) with respect to $t$. Hence, the specific form of interaction is used for mathematical simplicity in order to obtain analytical solutions. 
However, this form for the interaction term
it not physically very reasonable: it implies that the energy exchange
becomes zero when $H=0$. This property seems to be entirely unrealistic
since the energy exchange is mediated by local physical processes. There is
no reason why they should switch off if the expansion of the universe
momentarily halts everywhere at the expansion maximum of a closed universe,
or in large over-densities that reach an expansion maximum before collapsing
under their local self-gravity. Therefore, in what follows, we will not make
the usual (unrealistic) assumption that $Q\propto H$. Instead, we propose a
new interaction term $Q$ of the form 
\begin{equation}
Q=\frac{1}{t}(-\beta \rho _{1}+\alpha \rho _{2}),  \label{interaction}
\end{equation}

\noindent where $t$ is the cosmic time.  The interaction term is a leading order in a series expansion of possible contributions and
 allows us to build up intuition about the consequences of this type of interaction.  It is shown in \cite{PhD} that the interaction function is generally dependent on the scale factor under the form $Q\sim a^{-3}$. Consequently, the new interaction term is consistent with  respect to  dimensional analysis. 
The energy-exchange parameters, are
taken to be $\alpha $ and $\beta $, are real and positive, without loss of
generality.

A comment regarding the magnitude of the energy-exchange parameters $\alpha $
and $\beta $ is in order. If $\alpha =\beta $, then the strength of coupling
to the geometry of the 3-slice is taken to be the same for both fluids. In
general, the interaction term (\ref{interaction}) describes a `one-way'
interaction where the first fluid $\rho _{1}$ decays, whereas the second
fluid is gaining energy equal to that lost by the first fluid. This occurs
at early times where $\rho _{1}>\rho _{2}$. Then, at late times ($\rho
_{2}>\rho _{1}$), the interaction term changes sign as the second fluid now
decays and the first fluid gains energy. Generally, at early times, a
`one-way' interaction is feasible if $\frac{1}{t}(-\beta \rho _{1}+\alpha
\rho _{2})<0$ with $\rho _{1}>\rho _{2}$. Similarly, at late times, if $%
\frac{1}{t}(-\beta \rho _{1}+\alpha \rho _{2})>0,$ where $\rho _{2}>\rho
_{1} $.

In the early universe it is expected that density perturbations in the
matter give rise to the overdensities needed for the formation of galaxies 
\cite{Galaxy}. Therefore, by assuming an interaction between, say, dark
matter ($\rho _{1}$) and dark energy ($\rho _{2}$), it is expected that the
interaction term will be negative since $\rho _{1}>\rho _{2}$. Now, as the
expansion of the universe changes from decelerated to accelerated, the sign 
\cite{sign1,sign2,sign3} of the interaction term (\ref{interaction}) will
become positive in a natural way since $\rho _{2}>\rho _{1}$. We emphasize
here that our new interaction term (\ref{interaction}) is consistent with
the second law of thermodynamics \cite{sign2}, 
\begin{equation}
\dot{S_{1}}+\dot{S_{2}}=\left( \frac{1}{T_{1}}-\frac{1}{T_{2}}\right) Q\geq
0,  \label{entropy}
\end{equation}

\noindent where $\dot{S_{i}}$ represents the change of entropy of each fluid
in time, $T_{i}$ is the temperature for each fluid and $Q$ the energy
exchange. We expect the entropy of the universe to be non-decreasing with
time but we do not yet know all the repositories of entropy in the Universe.
Equation (\ref{entropy}) implies that as the universe expands, the
temperature of $\rho _{1}$ (assume dark matter) decreases while the
temperature of $\rho _{2}$ (assume dark energy) increases. Since at early
times $\rho _{1}>\rho _{2}$ (that is $Q<0$) and $T_{1}>T_{2}$, the total entropy of the system is positive. In addition, the law (\ref%
{entropy}) is satisfied at late times and the universe undergoes an
accelerated expansion, since $\rho _{2}>\rho _{1}$ (that is $Q>0$) and $%
T_{2}>T_{1}$. 

\section{The master equation}

\label{Section3} In this section we derive a master equation that governs
the evolution of the Hubble parameter $H$ for a system of two mutually
interacting fluids in a flat FRW universe, by means of the Friedmann
equation (units are chosen where $8\pi G=c=1$) 
\begin{equation}
3H^{2}=\rho _{1}+\rho _{2},  \label{friedmann}
\end{equation}

\noindent and the continuity equations for both fluids 
\begin{eqnarray}
\dot{\rho _{1}}+3H\Gamma \rho _{1} &=&\frac{1}{t}(-\beta \rho _{1}+\alpha
\rho _{2}),  \label{con2} \\
\dot{\rho _{2}}+3H\gamma \rho _{2} &=&\frac{1}{t}(\beta \rho _{1}-\alpha
\rho _{2}),
\end{eqnarray}

\noindent respectively. Using the last three equations, we can eliminate the
energy densities to obtain a single non-autonomous master equation for the
Hubble expansion 
\begin{equation}
\ddot{H}+\dot{H}\left[ \frac{1}{t}(\alpha +\beta )+3H(\Gamma +\gamma )\right]
+\frac{3H^{2}}{2t}(\alpha \Gamma +\beta \gamma )+\frac{9}{2}H^{3}\gamma
\Gamma =0.  \label{master}
\end{equation}

\noindent Under the transformation $H=u/t$ where $t=e^{x}$, the equation
above reduces to an autonomous, second-order non-linear differential equation
of the form

\begin{equation}
u{\prime \prime}+Au{\prime}+Buu{\prime}-(A+1)u+(C-B)u^2+Du^3=0,
\label{masteru}
\end{equation}

\noindent where prime denotes differentiation with respect to $x$ and 
\begin{equation}  \label{parameters}
A=(\alpha+\beta-3),\quad B=3(\Gamma+\gamma),\quad C=\frac{3}{2}%
(\beta\gamma+\alpha\Gamma),\quad D=\frac{9}{2}\Gamma\gamma.
\end{equation}

\noindent Equation (\ref{masteru}) is considered to be a Li\'{e}nard type
ordinary non-linear differential equation \cite{1,2}. In what follows, we
rewrite Eq. (\ref{masteru}) in the compact form 
\begin{equation}
u^{\prime \prime }+f(u)u^{\prime }+g(u)=0,  \label{abelu}
\end{equation}

\noindent where $f(u)=A+Bu$ and $g(u)=-(A+1)u+(C-B)u^2+Du^3$. It is argued
in \cite{1,2} that if the function $g(u)$ satisfies the condition 
\begin{equation}  \label{condition}
g(u)=f(u)\left[c_1+k\int{f(u)du}\right],
\end{equation}

\noindent where $c_{1}$ and $k$ are arbitrary constants, then Eq. (\ref%
{abelu}) is exactly integrable. After some manipulations, we find that this
integrability condition is satisfied for different pair of values for the
constants $c_{1}$ and $k$, as well as the parameters $A,B,C,$ and $D$. 

\subsection{Case I}

For the first case, condition (\ref{condition}) is satisfied for 
\begin{equation}  \label{cond1}
c_1=0,\quad k=-\frac{A+1}{A^2},\quad C-B=\frac{3}{2}kAB,\quad D=\frac{k}{2}%
B^2.
\end{equation}

\noindent The function $f(u)$ is given by $f(u)=A+Bu$, where the constants $%
A,B$ are defined above. Then, from the integrability condition (\ref%
{condition}), we obtain the function $g(u)$ as 
\begin{equation}
g(u)=kA^{2}u+\frac{3}{2}kBu^{2}+\frac{1}{2}kB^{2}u^{3}.
\end{equation}

\noindent The exactly integrable Li\'{e}nard equation is now given by 
\begin{equation}
u^{\prime \prime }+(A+Bu)u^{\prime }+\left( kA^{2}u+\frac{3}{2}kBu^{2}+\frac{%
1}{2}kB^{2}u^{3}\right) =0.  \label{abelu2}
\end{equation}

\noindent The equation above with $f(u)=A(u)+B(u)u$ was first used by
Dumortier and Rousseau\footnote{%
Where $A(u)$ a cubic function and $B(u)$ a linear function} in \cite%
{Dumortier} to describe the case of linearly forced isotropic turbulence.
Following the analysis expounded in \cite{1,2}, the non-linear equation (\ref%
{abelu2}) has a general solution in a parametric form and satisfies the
equation 
\begin{equation}
\frac{g(u)}{f(u)}=\tilde{C}^{-1}\exp [F(w,k)],  \label{quadratic}
\end{equation}

\noindent where $\tilde{C}^{-1}$ is a constant of integration and $\exp
[F(w,k)]$ is defined in \cite{1}. The time dependence of $u$ is determined
as a function of $w$ by 
\begin{equation}
x-x_{0}=\int \frac{dw}{f(u(w))(w^{2}+w+k)}.
\end{equation}

\noindent We note that Eq. (\ref{quadratic}) is equivalent to the quadratic
form 
\begin{equation}
Au+\frac{B}{2}u^{2}=\frac{\tilde{C}^{-1}}{k}\exp [F(w,k)],  \label{qq}
\end{equation}

\noindent and has real solutions for $k<1/4$. An exact solution of the
equation above is 
\begin{equation}
u(w)=\frac{-A\pm \sqrt{A^{2}+2B\tilde{C}^{-1}\exp [F(w,k)]/k}}{B},
\end{equation}

\noindent where $\exp [F(w,k)]$ has a special form for $k<1/4$ (see Eqn.
(17) in \cite{1}).

The exact parametric solutions also allow us to obtain some approximate
solutions of the differential equation (\ref{abelu2}). Assuming $w<<k$
(early time asymptote), Eqn. (\ref{qq}) takes the form 
\begin{equation}
\frac{B}{2}u^{2}\approx \frac{\tilde{C}^{-1}w}{k},
\end{equation}

\noindent and the parametric time evolution satisfies 
\begin{equation}
x-x_{0}\approx \frac{1}{k}\int {\frac{dw}{f(u(w))}}.
\end{equation}

\noindent After some manipulations we find that the asymptotic solution at
early times is 
\begin{equation}
u^{2}=\frac{2}{k}(\exp {[B\tilde{C}^{-1}(x-x_{0})]}-A^{2}),
\end{equation}

\noindent which gives time-evolution of the scale factor: 
\begin{equation}
a(t)\sim \exp {\left( t^{B\tilde{C}^{-1}/2}\right) }.  \label{scale1}
\end{equation}

\noindent The specific form of Eqn. (\ref{scale1}) describes a
time-evolution of the universe, with slower rate of expansion than the de
Sitter universe \cite{p3,24,25} called intermediate inflation \cite%
{inter1,inter} when $B\tilde{C}^{-1}/2<1$. If $B\tilde{C}^{-1}/2>2$ then
there would be a future curvature singularity. We note here that both fluids
dictate the asymptotic solution since $B=3(\Gamma +\gamma )$.

In the limit of large $w$, so that $w>>k$ and $w^{2}>>w$ (at late times), we
get the asymptotic form $\exp [F(w,k)]\approx \exp {(-k/(2w^{2}))}$, and 
\begin{equation}
\frac{B}{2}u^{2}\approx \frac{\tilde{C^{-1}}}{k}\exp {\left( -\frac{k}{2w^{2}%
}\right) }.
\end{equation}

\noindent The parametric time-evolution is 
\begin{equation}
x-x_{0}\approx \int {\frac{dw}{f(u(w))}w^{2}}.
\end{equation}

\noindent Following the same steps as above, we get the asymptotic solution 
\begin{equation}
\frac{B}{2}u^{2}=\frac{\tilde{C^{-1}}}{k}\exp {\left[ \frac{x(x-x_{0})^{2}}{%
2(A^{2}+2B\tilde{C^{-1}/k})}\right] },
\end{equation}

\noindent or, in terms of the scale factor, 
\begin{equation}
a(t)\sim \exp {\left( t^{\xi }\right) },  \label{scale2}
\end{equation}

\noindent where 
\begin{equation}
\xi=\frac{k}{A^2+2B/(\tilde{C}k)}.
\end{equation}

\noindent The asymptotic form of the solution (\ref{scale2}) is of interest
since it can describe the case of decelerated expansion throughout
cosmological evolution (since $k<1/4$). 

\subsection{Case II}

For the second case, condition (\ref{condition}) is satisfied for 
\begin{equation}  \label{case2}
c_1=-\frac{1}{B},\quad k=\frac{2D}{B^2},\quad A=0,\quad C=B.
\end{equation}

\noindent The equation above imposes some physical restrictions on the
energy exchange parameters and well as the barotropic indices of the fluids.
Specifically, for $A=0$ it occurs that $\alpha +\beta =3$, hence the
strength of coupling for each fluid with the geometry of the $3$-slice is
bounded. Also, for $C=B,$ we find that $\alpha =(2\Gamma -\gamma )/(\Gamma
-\gamma )$. Since $\alpha $ is taken to be positive the barotropic indices
must satisfy the condition 
\begin{equation}
\Gamma >\gamma ,\quad \text{or}\quad \Gamma <\frac{1}{2}\gamma .
\end{equation}

\noindent Assuming that the second fluid satisfies $\gamma =0$, then second
inequality opens up the possibility for the first interacting fluid being a
'phantom' \cite{phantom1,phantom2} satisfying $\Gamma <0$.

To find the exact solutions for the scale factor, we follow a different
approach to that for Case I. Taking into consideration the analysis used in 
\cite{chim}, we conclude that the differential equation 
\begin{equation}
u^{\prime \prime}+Buu^{\prime}-u+Du^3=0.  \label{abelu3}
\end{equation}

\noindent has the following invariant form, 
\begin{equation}
\bar{u}^{\prime \prime }+\bar{u}^{\prime }+\frac{2\bar{D}}{\bar{B}^{2}}\bar{u%
}-\frac{1}{\bar{B}}=0,  \label{invariant}
\end{equation}

\noindent where $\bar{u}=\int f(u)du$ and $\bar{x}=\int f(u)dx$. For the
homogeneous part of the linear equation above, the roots of the
characteristic equation are 
\begin{equation}
\lambda _{1,2}=\frac{-1\pm \sqrt{1-4\bar{k}}}{2},
\end{equation}

\noindent with $\bar{k}=2\bar{D}/\bar{B^2}$. A real solution of the form 
\begin{equation}
\bar{u}(\bar{x})=c_1\exp({\lambda_1 \bar{x}})+c_2\exp({\lambda_2 \bar{x}})+%
\frac{\bar{B}}{2\bar{D}},
\end{equation}

\noindent is obtained for $\bar{k}<1/4$. From (\ref{case2}) it follows that
the solution above describes the case of decaying fluids with $\Gamma
>\gamma $. We note here that for $\bar{k}=1/4$ the problem is reduced to
one-fluid description with no interaction since then $\Gamma =\gamma $.

The transformation of variables $\bar{u}=\int f(u)du$ and $\bar{x}=\int
f(u)dx$ relates the general solution of Eq. (\ref{abelu}) to Eq. (\ref%
{invariant}) through the transformation $\bar{y}(\bar{x})$. After some
calculations, we find that the general solution of (\ref{abelu}) has the
form 
\begin{equation}
u^{2}(x)=a^{-B/2}\left[ c_{3}a^{\sqrt{B^{2}-8D}/2}+c_{4}a^{-\sqrt{B^{2}-8D}%
/2}\right] +\frac{1}{D},  \label{u2}
\end{equation}

\noindent where $a$ is the scale factor and $c_{3},c_{4}$ are constants. For
simplicity, let us assume that the constant term in the equation above is
negligible. It follows that at early times, as $a\rightarrow 0$, we then
have 
\begin{equation}
u^{2}(x)\rightarrow a^{-(B+\sqrt{B^{2}-8D})/2},
\end{equation}

\noindent whereas, at late times where $a\rightarrow \infty $, 
\begin{equation}
u^{2}(x)\rightarrow a^{(-B+\sqrt{B^{2}-8D})/2}.
\end{equation}

\noindent These two equations can be integrated to obtain the following
exact solution for the scale factor 
\begin{equation}
a_{\pm }\propto (lnt)^{4/(B\pm \sqrt{B^{2}-8D})}  \label{scalefactor}
\end{equation}

\noindent at early ($+$branch) and late times ($-$branch) respectively. Now,
inserting (\ref{parameters}) into the latter solution we get the simplified
forms 
\begin{equation}
a\propto (lnt)^{2/(3\Gamma )},\text{ as }a\rightarrow 0,
\label{scalefactor1}
\end{equation}

\noindent and 
\begin{equation}  \label{scalefactor2}
a\propto (lnt)^{2/(3\gamma)},\text{ as }a\rightarrow\infty.
\end{equation}

\noindent The specific form of time evolution for the scale factor $%
a=(lnt)^{\xi }$ was previously found as one of the seven ever-expanding
possibilities in \cite{varietes}. Here, this specific type of solution
describes an expanding but non-inflationary universe for $\xi >0$.

The dynamic behaviours of the solutions (\ref{scalefactor1}) and (\ref%
{scalefactor2}) are determined by the barotropic indices $\Gamma $ and $%
\gamma ,$ respectively, and two types of evolution occur. First, assuming
that both barotropic indices are positive, the universe will experience an
initial singularity at a finite time $t^{\ast }$ and continue expanding. For
the second and more interesting scenario, assuming that the barotropic
indices have opposite signs \cite{chimlaz} (without loss of generality
assume $\Gamma >0$ and $\gamma <0$) then as $t\rightarrow \infty $, an
inflationary expansion will occur. Specifically, the universe experiences a
singularity at early times, followed by a period of decelerated expansion
(since $\xi >0$), and at late times the dominance of phantom fluid (with $%
\xi <0$) triggers the transition to an accelerated expansion.

In recent works, \cite{bulk1, bulk2}, the case of two interacting fluids in
the dark sector is modelled by a unified model of a single bulk viscous
fluid. There, the authors describe a transition from a phase of decelerated
expansion to one of accelerated expansion under some restrictions on the
value of bulk viscosity coefficient. 

\section{Curved case scenario}

\label{Section4} In this section we will study the case of curved FRW
universes filled with two interacting fluids. Our main purpose is to examine
how the evolution of the scale factor and Hubble parameter respectively are
changed, when we pass on to curved universes while keeping the interaction
term (\ref{interaction}) the same. Recall, that we were motivated in our
choice of interaction terms, Eq. (\ref{interaction}) by a desire to avoid
the explicit presence of $H$ in $Q$ in order to avoid unphysical behaviour
when $H=0$ and $H<0$ when an expansion maximum occurs. The Friedman equation
and conservation equations now read, 
\begin{eqnarray}  \label{con1}
3H^{2}+\frac{3k}{a^{2}} &=&\rho _{1}+\rho _{2},  \label{c3} \\
\dot{\rho _{1}}+3H\Gamma \rho _{1} &=&Q, \\
\dot{\rho _{2}}+3H\gamma \rho _{2} &=&-Q,
\end{eqnarray}

\noindent where the curvature parameter $k=\pm 1$, $a$ is the scale factor
and the interaction term still has the form $Q=\frac{1}{t}(-\beta \rho
_{1}+\alpha \rho _{2})$.

Now, we seek exact solutions for the scale factor. First, we take the time
derivative of Eqn. (\ref{friedmann}) and after manipulations we find the
evolution equation for the Hubble parameter $H(t)$ is, 
\begin{equation}
\dot{H}+\frac{3}{2}\gamma H^{2}=\frac{1}{2}(\gamma -\Gamma )\frac{m}{%
a^{3\Gamma }}+\frac{k}{a^{2}}(1-3\gamma /2).  \label{evolutionH}
\end{equation}

\noindent Here, we have used the relation \cite{up} 
\begin{equation}
\rho _{1}=\frac{\int {Qa^{3\Gamma }}dt}{a^{3\Gamma }}\equiv \frac{m}{%
a^{3\Gamma }},
\end{equation}

\noindent which occurs after integrating Eqn. (\ref{con1}), and

\begin{equation*}
m(t)\equiv \int {Qa^{3\Gamma }}dt.
\end{equation*}

In terms of the scale factor the evolution equation (\ref{evolutionH}) reads 
\begin{equation}
\frac{\ddot{a}}{a}-\frac{1}{a^{2}}(1-3\gamma /2)(k+\dot{a}^{2})=\frac{1}{2}%
(\gamma -\Gamma )\frac{m(t)}{a^{3\Gamma }}.
\end{equation}

\noindent Using the conformal time transformation $ad\eta =dt$ and new
variable \cite{observations} $b=a^{3\Gamma -2}$, it can be shown that the
equation above is equivalent to 
\begin{equation}
b^{\prime \prime }+\frac{k}{2}(2-3\gamma )(2-3\Gamma )b=\frac{1}{2}%
(2-3\Gamma )(\Gamma -\gamma )m(\eta ),  \label{forced}
\end{equation}

\noindent where $^{\prime }\equiv d/d\eta $, which in turn describes the
equation of motion for a forced harmonic oscillator subject to a
time-dependent force for $Q\not=0$. We note here that this equation is valid
under the constraint $6\Gamma =2+3\gamma $, which restricts the possible
type of cosmological fluids used.

The solution to the associated homogeneous equation reads 
\begin{equation}
b_{c}(\eta )=c_{1}\cos {[\sqrt{(2-3\Gamma )(2-3\gamma )k/2}\eta ]}+c_{2}\sin 
{[\sqrt{(2-3\Gamma )(2-3\gamma )k/2}\eta ]},  \label{bc}
\end{equation}

\noindent where $\omega _{0}^{2}=(2-3\gamma )(2-3\Gamma )k/2$ is the natural
frequency of oscillations and depends on the barotropic indices of the
interacting fluids and the spatial curvature. Any general time-dependent
periodic function can be decomposed into its Fourier cosine and sine
components. Assuming that the external force $F_{e}(\eta )$ given by the
right-hand side of Eqn. (\ref{forced}) is periodic, the particular solution
of the associated problem reads, 
\begin{eqnarray*}
b_{p}(\eta ) &=&\frac{\gamma -\Gamma }{k(3\gamma -2)}\cos {[\sqrt{(2-3\Gamma
)(2-3\gamma )k/2}\eta ]}\int {m(\eta )\sin {[\sqrt{(2-3\Gamma )(2-3\gamma
)k/2}\eta ]}d\eta }  \label{bp} \\
&+&\frac{\gamma -\Gamma }{k(3\gamma -2)}\sin {[\sqrt{(2-3\Gamma )(2-3\gamma
)k/2}\eta ]}\int {m(\eta )\cos {[\sqrt{(2-3\Gamma )(2-3\gamma )k/2}\eta ]}%
d\eta }.  \notag
\end{eqnarray*}

\noindent The general solution to Eqn. (\ref{forced}) is a combination of
the solution for the homogeneous part of the equation, $b_{c}$ and the
particular solution, $b_{p}$. Thus, 
\begin{eqnarray*}
b(\eta ) &=&c_{1}\cos {[\sqrt{(2-3\Gamma )(2-3\gamma )k/2}\eta ]}+c_{2}\sin {%
[\sqrt{(2-3\Gamma )(2-3\gamma )k/2}\eta ]} \\
&+&\frac{\gamma -\Gamma }{k(3\gamma -2)}\cos {[\sqrt{(2-3\Gamma )(2-3\gamma
)k/2}\eta ]}\int {m(\eta )\sin {[\sqrt{(2-3\Gamma )(2-3\gamma )k/2}\eta ]}%
d\eta } \\
&+&\frac{\gamma -\Gamma }{k(3\gamma -2)}\sin {[\sqrt{(2-3\Gamma )(2-3\gamma
)k/2}\eta ]}\int {m(\eta )\cos {[\sqrt{(2-3\Gamma )(2-3\gamma )k/2}\eta ]}%
d\eta }.
\end{eqnarray*}

\noindent For simplicity, we assume an \textquotedblleft
one-way\textquotedblright\ interaction of the form $Q=-\beta \rho _{1}t^{-1}$%
\footnote{%
Here, the interaction term is similar to the one used in \cite{up}, if we
assume that the scale factor describes a Milne universe so that $a=t$.} so
that the first fluid decays into the second fluid at a rate proportional to
its energy density. As a result, the energy loss from the first fluid is
gained by the second fluid and the total energy is conserved. It is shown
(see Appendix in \cite{up}) that for such interaction term, the
corresponding time-dependent function $m$ takes the form $m(\eta
)=m_{0}e^{-\beta \eta }$, where $m_{0}$ a constant of integration. The
solution for $k\not=0$ then takes the form 
\begin{equation}
b(\eta )=c_{1}\cos {[\sqrt{(2-3\Gamma )(2-3\gamma )k/2}(\eta -\eta _{0})]}+%
\frac{(\gamma -\Gamma )m_{0}e^{-\beta \eta }}{k(3\gamma -2)+\beta ^{2}}.
\label{b1way}
\end{equation}

\noindent

\subsection{An Oscillating toy-model}

Let us now examine the behavior of the exact solution found above for a
positively curved universe containing radiation ($\Gamma =4/3$) and scalar
field ($\gamma =2$). Both barotropic indices satisfy the constraint imposed
when obtaining Eqn. (\ref{forced}). We assume without loss of generality
that the energy exchange parameter $\beta $ is negative so that energy is
being transferred from scalar field (that is $\rho _{2}$) to radiation ($%
\rho _{1}$). The exact solution in terms of the scale factor is, 
\begin{equation}
a^{2}(\eta )=|c_{1}\cos {[2(\eta -\eta _{0})]}|+\frac{2m_{0}e^{\beta \eta }}{%
3(4+\beta ^{2})}.  \label{osc}
\end{equation}

\noindent We use this form to avoid negative values of the scale factor and
ensure that it will be real and semi-finite throughout cosmic evolution.

At early times, the oscillatory part of the solution dominates the expansion
of the universe. Each successive maximum is separated by a collapse of the
scale factor to zero size. Energy is being transferred from the scalar field
to the radiation at a slow rate which results in the same amplitude of
oscillation for each cycle. As time increases, the amplitude of each cycle
increases monotonically and asymptotically the universe is pushed closer and
closer to spatial flatness, with each successive cycle being longer and
longer lived \cite{PhD, 28, gang}.

Moreover, the form of solution (\ref{osc}) suggests that after a number of
oscillatory cycles, a collapse singularity will be avoided in the late
future. Rather, a series of non-zero expansion minima will take place. This
behavior occurs because at this point of evolution, the endless energy
transfer of energy from scalar field to radiation results in the former
obtaining negative energy density allowing non-zero expansion minima \cite%
{up, 28,Tolman}. At late times, the universe is expected to experience a
run-away behavior as the exponential term dominates the evolution. As a
result the universe will expands eternally. This is characteristic of the
behaviour of closed universes with a positive cosmological constant \cite{28}%
. Entropy increase will increase the size of the successive expansion maxima
until they become large enough for the cosmological constant to influence
the dynamics. When it does, the oscillations will cease and the expansion
will approach de Sitter expansion as $t\rightarrow \infty $. The final state
will be dominated by the cosmological constant but lie quite close to the
state where the density parameters of the cosmological constant and the
matter, with a closeness determined by the size of the entropy increase from
one cycle to the next. This evolution will change in some respects if
anisotropies are included. Although successive cycles increase in size and
entropy they also become increasingly anisotropic at their expansion volume
maxima. If a cosmological constant stops these oscillations it will
isotropise the expansion on approach to de Sitter as $t\rightarrow \infty ,$ 
\cite{gang, gang1}.

\section{Discussion}

In this work, we have investigated both the behavior of solutions of
spatially flat and curved FRW universes, with the inclusion of mutual energy
exchange between two cosmological fluids. In particular, the interaction
term we employ has a more realistic form than is generally employed in the
literature, ensuring that the energy exchange between the interacting fluids
does not cease when the Hubble parameter is zero.

The case of interacting fluids has attracted growing interest over the past
years, and the idea of energy exchange between cosmological fluids has been
espoused as a mean to explain recent observational data. To be more precise,
the presence of a coupling between dark energy and dark matter is an attempt
to alleviate the 'coincidence problem' between their present energy
densities \cite{lip}. In addition, various physical processes such as
particle-antiparticle annihilation into radiation, black hole evaporation,
particle decays, and vacuum decay into radiation can be modelled as examples
of interacting cosmic fluids.

Considering all these possible applications, we look for solutions for the
expansion of the scale factor of FRW universes of all curvatures and none,
in the presence of energy exchange between two fluid components using a
non-linear interaction term. First, we considered the case of a system of two
mutually interacting fluids in flat FRW universes. Using the Friedman
equation and the conservation equations for both fluids we obtained a
second-order, non-autonomous, non-linear differential equation describing the
evolution of the Hubble parameter. Under a transformation, the master
equation is reduced to a single autonomous, non-linear differential equation.
The reduced differential equation is identical to an exact, integrable Li%
\'{e}nard type equation under some restrictions. To find exact solutions to
the problem, we followed two different techniques expounded in the
literature.

For the first case, we use the Chiellini integrability condition to obtain a
class of exact solutions of the Li\'{e}nard equation expressed in a
parametric form. At early times we show that the asymptotic solution for the
scale factor satisfies describes a so-called phase of intermediate
inflation. On the other hand, at late times the form of the scale factor
shows that the universe undergoes decelerated expansion throughout cosmic
evolution.

For the second case, the reduced differential equation is written under a
form-invariant transformation as a second order, linear, non-homogeneous
differential equation. As before, two asymptotic solutions are obtained at
early and late times respectively. Both solutions found describe the case of
an expanding but non-inflationary universe (assuming both barotropic indices 
$\gamma ,\Gamma $ are positive). In the case, if we allow barotropic indices
with opposite signs, then the presence of phantom fluids drives the
evolution of the universe towards an accelerated expansion.

In the last section, the case of interacting fluids in curved FRW
cosmologies is examined. Under a conformal transformation and a restriction
on the barotropic indices, the master equation describing the expansion of
the scale factor is identical to the form of a forced harmonic oscillator
(for non-zero interaction term). We have shown that the general solution
governing the evolution of the universe in the case of positive spatial
curvature follows an eternal series of cycles at early times. Each one of
these cycles begins with a big bang singularity and ends with a big crunch
singularity. After each crunch-to-bang transition, the entropy of the
universe increases, As a result, the amplitude of oscillations increases
progressively. A late times, the presence of the exponential term dominates
the oscillatory expansion the universe and after a finite number of future
cycles there are no further oscillations. Thus, at late times the universe
is singular free. Finally, assuming a toy-model of one-way of interaction
between radiation and scalar field, we show that, as expected at late times,
the collapse singularity after each successive maximum is avoided.

\bigskip

\textbf{Acknowledgements:} JDB is supported by the Science and Technology
Facilities Council (STFC) of the UK.

\end{document}